# An Actuator with Magnetic Restoration, Part I: Electromechanical Model and Identification

Sajjad Mohammadi, *Member, IEEE*, William R. Benner, James L. Kirtley, *Life Fellow*,
Jeffrey H. Lang, *Life Fellow*

*Abstract*—electromechanical models are crucial in the design and control of motors and actuators. Modeling, identification, drive, and current control loop of a limited-rotation actuator with magnetic restoration is presented. New nonlinear and linearized electromechanical models are developed for the design of the drive as well as small and large signal controls of the actuator. To attain a higher accuracy and an efficient design, and the eddy-currents in the laminations and magnet are modeled. This involves analytically solving 1-D and 2-D diffusion equations, leading to the derivation of a lumped-element circuit for system-level analyses, such as control system design. Additionally, the study analyzes and incorporates the impact of pre-sliding friction. The actuator is prototyped, and the paper delves into the identification of the model, presenting a procedure for parameter extraction. A close agreement is observed between the results obtained from the model, finite element analysis, and experimental results. The superiority of the proposed model over previous approaches is highlighted. Part II of the paper is dedicated to the drive circuit, the current control, as well as linear and nonlinear position control system designs.

*Index Terms*—actuator, control, eddy current, diffusion, electric machines, motors, FEM, modeling

## I. Introduction

ROTARY actuators has become widespread across various industries, ranging from robotics and aerospace to fluid valves and optical scanning. These actuators offer several advantages, including a simple structure, cost-effective maintenance, high reliability, low cost, and simple control. When designed to provide a constant torque, they are referred to as limited-angle torque motors (LATMs) [1]-[3]. Another common type for such actuators is voice coil motors (VCMs) or voice coil actuators (VCAs), drawing from the historical use of the "voice coil" term in audio speakers. Despite its origin in audio systems, this technology has found applications beyond the realm of audio. In specific scenarios, such as fail-safe operations, there is a need for restoration torque. This can be achieved through the utilization of nonlinear stiffness, as seen in Laws's relays [4], or by incorporating alignment poles [5]. This paper presents generalized studies applicable to such actuators, with certain aspects of the physical implementations covered by patents (e.g., [6]-[7]).

Achieving high-performance control of electric machines necessitates precise models and efficient identification processes, which can be conducted either offline [8] or online [9]. Although the finite element method (FEM) stands out as a powerful numerical modeling technique [10], it is deemed too slow for dynamic studies. Magnetic equivalent circuits (MECs) [11]-[14] and subdomain models [15]-[16] offer rapid yet precise analytical models for design and dynamic analysis of electromechanical devices, ranging from motors and actuators to couplers and sensors. The influence of eddy currents on the frequency characteristics of inductors or the dynamic behavior of motors and actuators can be significant. Therefore, incorporating their effects into the models is crucial for more accurate simulations and more efficient control system designs.

Efforts have been made to model the impact of eddy currents in various electromagnetic devices. In [12]-[14], an analytical model is developed for eddy-current couplers with different geometries by combining MECs with Faraday's and Ampere's laws. Papers [15]-[16] present an analytical method for rotational and linear eddy-current-based speed sensors, using the solution of the 2-D diffusion equation in cylindrical coordinates and Bessel's equation. Valuable contributions to modeling eddy-current impacts in magnetic cores and inductors have also been made [17]-[19]. In [17], a MEC-based model is employed to obtain a 1-D eddy current solution across the lamination thickness of magnetic cores. In [18], a model for laminated iron-core inductors at high frequencies is developed by calculating 1-D eddy currents in the laminations using Ampere's law. An augmented circuit model for magnetic bearings is introduced [19], accounting for the effects of 1-D eddy currents in the core laminations as a parasitic winding that drives a ladder network of resistors and inductors.

Previous works in modeling actuators have some deficiencies, such as ignoring eddy currents in the iron and magnets, considering them only in the laminations, and neglecting the impact of friction. In [20], the impact of eddy-current damping on the operation of rotary actuators is studied. In [21], a MEC-based model for the design of limited-angle actuators is developed, where eddy currents and friction are ignored. In [22], the diffusion equation for the solution of eddy currents in non-laminated cores of a reluctance actuator is solved to develop a circuit model. In [23], a linear model is developed for a VCMs, where nonlinearities of coil and spring torques, and impacts of eddy currents are ignored. In [24], a MEC-based electromechanical model is developed for dynamic response optimization of an actuator, where eddy currents are ignored. In [25], a model is developed for rotary VCAs, where eddy currents are ignored, leading to some inaccuracies. In [26]-[29], various control systems for VCMs and actuators are implemented using electrotechnical models, in which eddy currents in laminations and magnets, as well as model nonlinearities are ignored. Friction, impacting the mechanical dynamics, can be studied using the LuGre model [30]-[31].

TABLE I
SPECIFICATIONS OF THE STUDIED MOTOR

| parameter | value | parameter | value |
|---|---|---|---|
| outer diameter, $D_o$ | 13.716 mm | PM remnant, $B_r$ | 1.37 Tesla |
| lamination thickness $d$ | 0.35 | total turns, $N$ | 100 |
| # of laminations, $m$ | 12 | wire gauge | AWG33 |
| stack length, $L$ | 4.191 mm | torque constant, $k_t$ | 1.906 mN.m/A |
| pole width, $w_p$ | 4.72 mm | Mag. spring $k_s$ | 0.636 mN/rad |
| PM length, $L_{pm}$ | 9 mm | total stiffness, $K_s$ | 1.3 mN/rad |
| rotor diameter, $D_r$ | 3.048 mm | total damping, $k_d$ | 4.49e-7 Ns/rad |
| minor radius, $R_1$ | 1.71 mm | inertia, $J$ | 1.65e-9 kg.m$^2$ |
| major radius, $R_2$ | 1.9665 mm | inductance, $L_{c0}$ | 280 uH |
| PM conductivity | 0.6 MS/m | resistance, $R_c$ | 1.76 ohm |
| iron conductivity | 2 MS/m | sense resistor, $R_s$ | 0.1 ohm |

In [32], the authors developed a new analytical model for the actuator studied in this paper, focusing on device design and understanding electromagnetic fundamentals. However, this model is computationally burdensome for dynamic studies. In the current paper, the authors take a step further by incorporating motion and unmodeled dynamics, creating am electromechanical model suitable for drive and control studies.

The main contribution of this paper is the development of a nonlinear electromechanical model for a rotary actuator with magnetic restoration, encompassing eddy-currents and friction. The analytical representation of the rotor incorporates lumped-element and Amperian current models for the magnet. The paper derives nonlinear and linearized relationships for coil torque, restoration torque, and back-emf. To enhance modeling precision, the eddy currents in the stator laminations are modeled using the analytical solution of the one-dimensional diffusion equation, while eddy currents in the magnets are calculated from the solution of the two-dimensional diffusion equation in Cartesian coordinates. To account for the impact of permeabilities of iron, air, and magnet, effective permeabilities are calculated, providing additional precision to the 1-D and 2-D diffusion equations. To facilitate dynamic studies, the analytical field solutions are transformed into lumped-element magnetic circuits by introducing frequency-dependent reluctances. Finally, the lumped models for the eddy currents in the laminations and magnet are combined and incorporated into the electromechanical model of the actuator by introducing a frequency-dependent inductance to the electrical dynamics, making it more suitable for system-level analysis and control system designs. Additionally, the paper analyzes the impact of pre-sliding friction, incorporating it into the stiffness and damping of the model. The developed nonlinear electromechanical model is versatile, suitable for large-signal control, while it is also linearized to facilitate linear control system designs for small-signal tracking. Furthermore, 2-D and 3-D FEM are employed in the analysis. The paper outlines a procedure for the identification and parameter extraction of the model, demonstrating a close agreement between the results obtained from experiments, the model, and FEM.

The upcoming Part II of the paper will showcase the model's capability in achieving near-zero discrepancy when estimating the phase margin of the current control loop. This precision is crucial and could have significant errors if eddy currents were neglected or only considered in the laminations as is the case for the existing models. The paper will proceed to discuss the prototyping of the actuator. Part II will delve into the design and modeling of a drive circuit and the current control loop. It will then analyze and experimentally evaluate the modeling accuracy, design effectiveness, and practical trade-offs. The conclusion will involve the implementation of linear and nonlinear position control systems for various applications.

## II. THE ACTUATOR

The geometry and the exploded view of the actuator, whose specifications are listed in Table I, are shown in Fig 1(a)-(b). The rotor PM has diametral magnetization. The interaction of the stator flux and the magnet produces the main torque with $sin\ \theta$ distribution and a peak at $\theta=90°$. The stator inner surface is shaped to have an elliptical curvature whose interaction with the magnet produces a reluctance torque with $sin\ 2\theta$ distribution and peaks at $\theta=45°$ and $\theta=135°$ as well as a stiffness which tends to restore the rotor back to the maximum torque per ampere position (MTPAP), i.e. $\theta=90°$. At small-angle rotations around $\theta=90°$, this stiffness acts like a linear spring. Such limited-angle rotary actuators have applications from robotics, mechatronics and medical devices to laser scanning, 3D printers and fluid valves.

## III. TORQUE, BACK-EMF, AND DESIGN CONSIDERATIONS

### A. Permanent Magnet Model

The magnetization vector $M$ of the PM in terms of azimuth $\varphi$ and rotor angular position $\beta$ can be represented as in below:

$$\vec{M}(\varphi,\beta) = -M\sin(\varphi-\beta)\ \hat{r} - M\cos(\varphi-\beta)\ \hat{\varphi};\ r \leq R_r \quad (1)$$

A magnetization $M$ can be represented as Amperian current density $J_m$. As shown in Fig. 1(c), since $M$ is uniform inside the PM, there is only a surface current density $K_m$ as

$$\begin{cases} \vec{J}_m = \nabla \times \vec{M} \\ \vec{K}_m = \vec{M} \times \hat{n} \end{cases} \quad (2)$$

where $n=r$ is the unit vector normal to the surface. Thus

$$\begin{cases} \vec{K}_m(r,\varphi,\beta) = \vec{M} \times \hat{r} \\ \vec{K}_m(r,\varphi,\beta) = M\cos(\varphi-\beta)\ \hat{z};\ r = R_r \end{cases} \quad (3)$$

As shown in Fig. 1(d), the lumped-element model of the PM consists of a permeance $\wp_m$ and a magneto-motive force $F_m$ which is the total current enclosed in the Amperian loop as

$$F_m = \int_{\varphi=-\pi/2+\beta}^{\varphi=\pi/2+\beta} |K_m(\varphi,\beta)| R_r d\varphi = 2R_r M \tag{4}$$

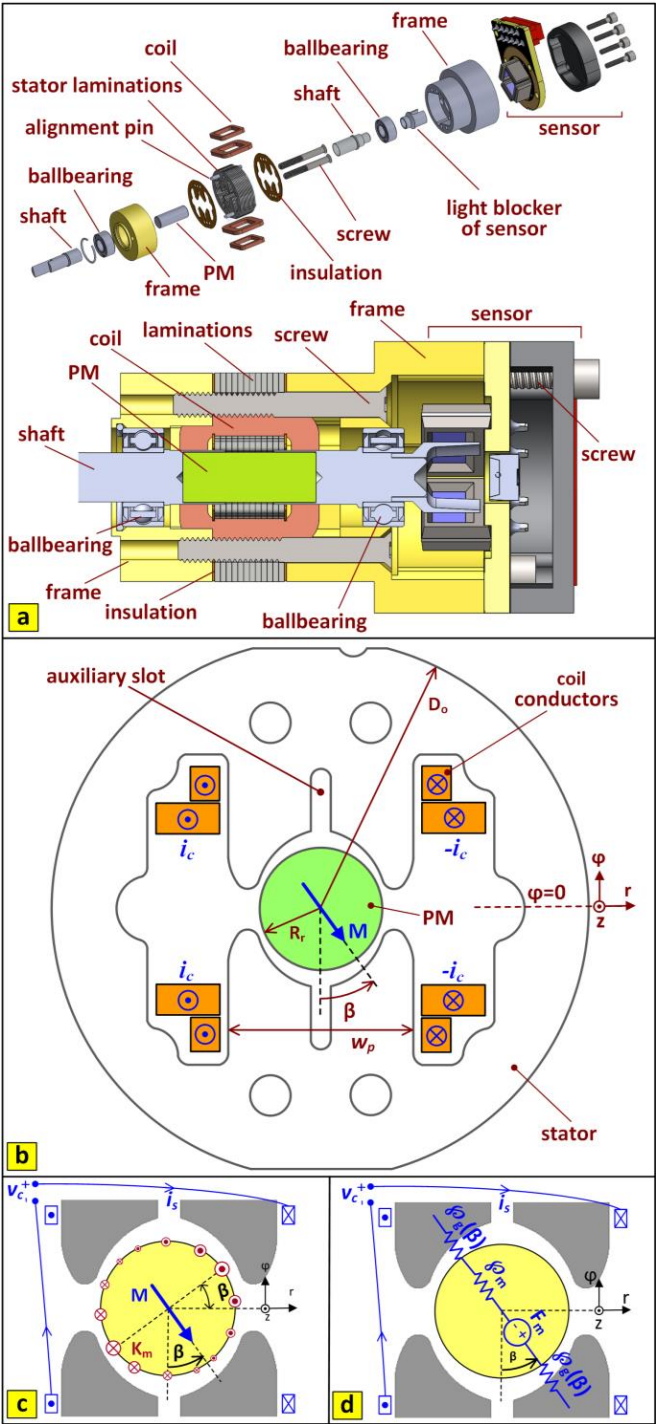

Fig. 1. (a) Exploded view of the actuator, (b) geometry of the actuator, (c) Amperian current model of PM, and (d) lumped-element models of the PM

### B. Stator Field

Through Ampere's law, the current in the stator coils produces a magnetic field as

$$\nabla \times \vec{B} = \mu_0 \mu_r \vec{J} \tag{5}$$

The radial component of magnetic flux density distribution on the surface of the PM, which is the torque-producing component, can be represented in Fourier series as

$$B_r(\varphi) = \sum_{n=1,3,5}^{+\infty} B_n \sin n\varphi = B_1 \sin \varphi + B_3 \sin 3\varphi + \ldots \tag{6}$$

As long as the stator iron is not saturated, the coefficients $B_n$ are linearly proportional to the coil current $i_c$, so

$$B_r(\varphi) = \sum_{n=1,3,5}^{+\infty} k_n i_c \sin n\varphi = k_1 i_c \sin \varphi + k_3 i_c \sin 3\varphi + \ldots \tag{7}$$

### C. Coil Torque

The stator flux interacts with the PM to produce an electromagnetic torque which is obtained by Lorentz force as

$$T_{coil} = L \int_0^{2\pi} R_r K_m(\varphi,\beta) B_r(\varphi) R_r d\varphi \tag{8}$$

By substitution of $K_m$ and $B_r$, we have:

$$T_{coil}(\beta, i_c) = L \int_0^{2\pi} R_r M_0 \cos(\varphi-\beta) \left( \sum_{n=1,3,5}^{+\infty} k_n i_c \sin n\varphi \right) R_r d\varphi \tag{9}$$

Except for $n=1$, the integration of the product of $\cos(\varphi-\beta)$ and $\sin n\varphi$ is zero, i.e., only the fundamental component of $B_r$ contributes to the torque production. It simplifies as

$$T_{coil}(\beta, i_c) = L R_r^2 M_0 k_1 i_c \int_0^{2\pi} \cos(\varphi-\beta) \sin(\varphi) d\varphi \tag{10}$$

By expressing the trigonometric product in sums, it yields

$$\begin{cases} T_{coil}(\beta, i_c) = k_t i_c \sin \beta \\ k_t = \pi L R_r^2 k_1 M_0 \end{cases} \tag{11}$$

where $k_t$ is the torque constant [Nm/A].

### D. Restoration Torque

As the PM is faced the maximum permeance at MTPAP ($\beta=90$), the total permeance can be expressed as

$$\wp(\beta) = \wp_0 - \wp_1 \cos 2\beta \tag{12}$$

The stored co-energy and restoration torque are obtained as

$$W_c(\theta) = \frac{1}{2} F_m^2 \wp(\beta) \tag{13}$$

$$\begin{cases} T_{rest} = \dfrac{\partial W_c(\beta)}{\partial \beta} = \dfrac{1}{2} F_m^2 \dfrac{\partial \wp(\beta)}{\partial \beta} \\ T_{rest} = k_{rest} \sin 2\beta; \quad k_{rest} = \wp_1 F_m^2 \end{cases} \tag{14}$$

where $k_{rest}$ is the maximum restoration torque.

### E. Total Torque

The total electromagnetic torque can be expressed as

$$T_e(\beta, i_c) = k_t i_c \sin \beta + k_{rest} \sin 2\beta \tag{15}$$

whose small-signal model around MTPAP ($\theta=\beta-\pi/2$) is

$$T_e(\theta, i_c) = k_t i_c - k_s \theta \tag{16}$$

where $k_s = 2 k_{rest}$ can be defined as the magnetic spring constant.

### F. Back Electromotive Force

The flux linked by the stator coil is as

$$\begin{cases} \lambda(\beta, i_c) = L_{co} i_c + \lambda_m(\beta) \\ \lambda_m(\beta) = -\lambda_0 \cos \beta \end{cases} \tag{17}$$

where $\lambda_m$ and $\lambda_0$ are PM flux and its maximum, and $L_{c0}$ is the frequency-independent coil inductance. As PM flux is in the opposite direction of the unit normal vector of coil area at $\beta=0$, there is a negative sign. The back-emf is obtained as

$$\begin{cases} E(\omega_r, \beta) = \dfrac{d\lambda_m}{dt} = \dfrac{d\lambda_m}{d\beta}\dfrac{d\beta}{dt} = \omega_r \dfrac{d\lambda_m}{d\beta} \\ E(\omega_r, \beta) = \lambda_0 \omega_r \sin\beta \end{cases} \quad (18)$$

where $\beta=\omega_r t$ and $\omega_r=d\beta/dt$ is the angular velocity. Defining the back EMF constant $k_b$ as the amplitude at 1 *rad/sec*, we have

$$E(\omega_r, \beta) = k_b \omega_r \sin\beta \quad (19)$$

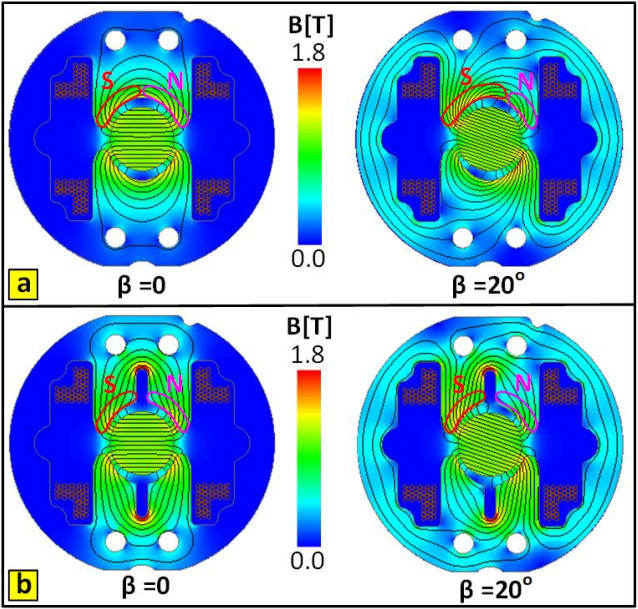

Fig. 2. PM flux and hysteresis effect: (a) without and (b) with auxiliary slots.

In the linearized model around MTPAP, $E=k_b \omega_r$. Due to energy conservation in the conversion of electrical power ($E i_s$) to mechanical form ($T_{coil}\omega_r$), $k_b=k_t$.

*G. Design Considerations and Flux Analysis*

The rotor radius $R_r$ and thus the overall sizing is obtained based on torque/power requirements. The inner radius of the stator is designed to provide the winding space according to the required electrical loading. The outer diameter of the stator $D_o$ is designed such that the back iron operates at the knee point of the magnetic saturation curve; too small values result in saturation while a large value causes the excessive use of iron and oversizing. There is a compromise between $k_t$ and $k_{res}$; a larger restoration can be achieved by a higher saliency, but $k_t$ goes down as the saliency increase the effective air-gap length.

As illustrated in Fig. 2, there are two auxiliary slots to divide the pole faces into two sections to suppress hysteresis effects. As the rotor oscillates around MTPAP, without the auxiliary slots, the direction of the PM flux pro duced within each half of the pole faces changes, so there could be a hysteresis effect making one-half of the pole face more or less North/South if the current is removed when the rotor is not at the MTPAP; as a result, the rotor position is restored with small deviation from MTPAP. By separating the two halves of a pole face, the magnet flux turns the auxiliary slot; thus, one section always stays North and the other one always stays South. The opening of these two slots should be small enough so that its fringing

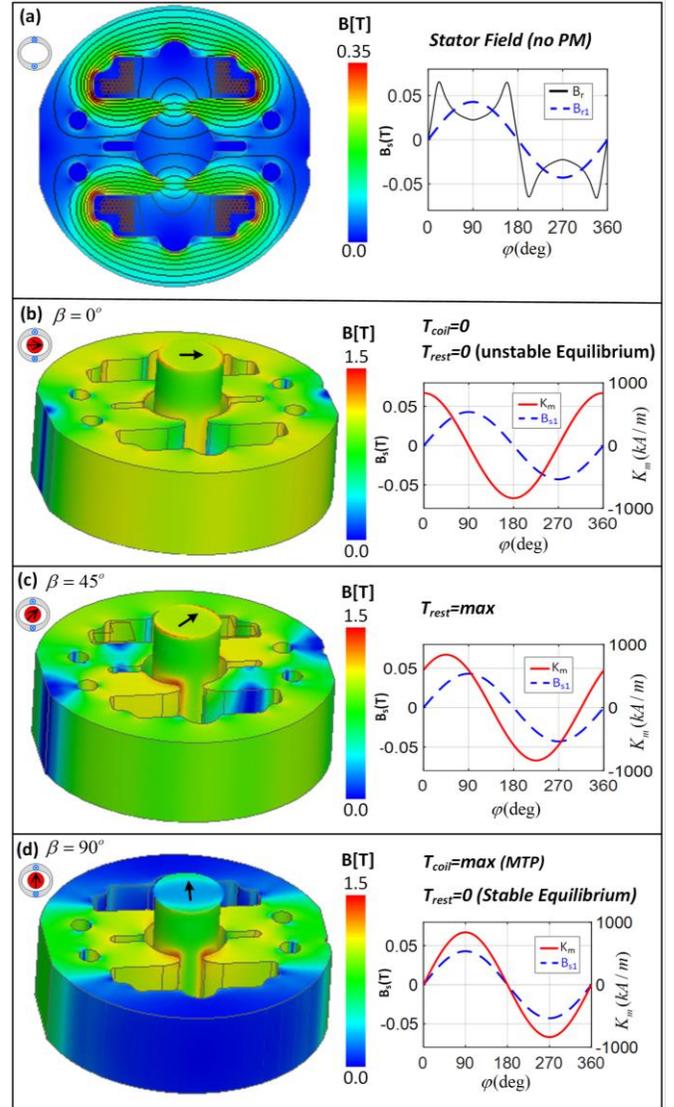

Fig. 3. (a) 2D distribution of magnetic flux density and flux lines (left), and radial component of magnetic flux density $B_r$ and its fundamental $B_{r1}$ due to stator current of *1A*, and (b)-(d) 3D distribution of magnetic flux density (left), and Amperian current distribution of PM together with $B_{r1}$ (right) at rotor positions $\beta=0$, $\beta=45^o$ and $\beta=90^o$.

effect can be ignored.

Fig. 3(a) shows flux lines, flux density distribution, the radial component $B_r$ and its fundamental $B_{r1}$ on the rotor surface due to the coil current. Figs. 3(b)-(d) show the flux density distribution due to the PM at different rotor positions as well as the PM Amperian currents $K_m$ together with $B_{r1}$—the torque producing components. At $\beta=0$, $K_m B_{r1}$ integrates to zero, so $T_{coil}=0$; also, $T_{rest}=0$, because the PM is faced with the minimum permeance, which is an unstable equilibrium as the slope of the curve is positive. At $\beta=45$, $T_{rest}$ is maximum. At MTPAP, i.e., $\beta=90$, $K_m B_{r1}$ integrates to a maximum value; also, $T_{rest}=0$ as the PM is faced with the maximum permeance, which is a stable equilibrium as the slope of the curve is negative.

## IV. ELECTROMECHANICAL MODEL

*A. Nonlinear Electromechanical Model*

The governing electromechanical dynamic, whose block

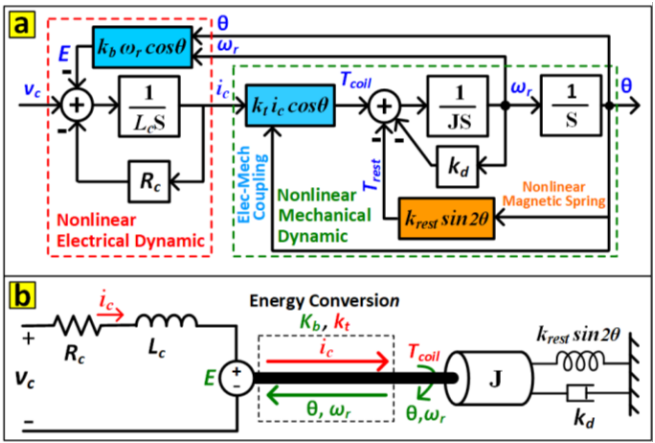

Fig. 4. The developed nonlinear electromechanical model: (a) block diagram, and (b) electrical and mechanical elements

diagram and elements are shown in Fig. 4, is as follows

$$v_c(t) = R_c i_c(t) + \frac{d\lambda(\beta, i_c)}{dt} \Rightarrow$$

$$v_c(t) = E(t) + R_c i_c(t) + L_{co} \frac{di_c(t)}{dt} \quad (20)$$

$$J\frac{d^2\beta}{dt^2} + k_d \frac{d\beta}{dt} = T_e(\beta, i_c) - T_L \quad (21)$$

where $k_d$ is the viscous damping constant, and $T_L$ is the load torque. It leads to a nonlinear differential equation as

$$J\ddot{\beta} + k_d \dot{\beta} - k_{rest} \sin 2\beta = k_t i_c \sin \beta - T_L \quad (22)$$

The states are defined as angular position, angular velocity, and coil current. The inputs are coil voltage and load torque as

$$x(t) = \begin{bmatrix} x_1 \\ x_2 \\ x_3 \end{bmatrix} = \begin{bmatrix} \beta \\ \omega_r \\ i_c \end{bmatrix} \; ; \; u(t) = \begin{bmatrix} u_1 \\ u_2 \end{bmatrix} = \begin{bmatrix} v_c \\ T_L \end{bmatrix} \quad (23)$$

Substitution for $E$ and $T_t$ yields the nonlinear system below

$$\begin{cases} \dot{x}_1 = x_2 = f_1 \\ \dot{x}_2 = \dfrac{-k_d x_2 + k_t x_3 \sin x_1 + k_{rest} \sin 2x_1 - T_L}{J} = f_2 \\ \dot{x}_3 = \dfrac{-R_c x_3 - k_t x_2 \sin x_1 + v_c}{L_{co}} = f_3 \end{cases} \quad (24)$$

### B. Equilibrium Point

The equilibrium points, i.e., the solution of the system of equation $[f_1=0; f_2=0; f_3=0]$ at zero input, are obtained as

$$\begin{cases} \bar{\beta} = 0, \pi/2, \pi, 3\pi/2 \\ \bar{\omega}_r = 0 \\ \bar{i}_c = 0 \end{cases} \quad (25)$$

where $\pi/2$ and $3\pi/2$ are stable equilibriums, and $0$ and $\pi$ are unstable ones. The position $\beta=\pi/2$ is taken as MTPAP.

### C. Linearized Electromechanical and State Space Models

The system is linearized around the equilibrium point below

$$\bar{x} = \begin{bmatrix} \bar{x}_1 \\ \bar{x}_2 \\ \bar{x}_3 \end{bmatrix} = \begin{bmatrix} \pi/2 \\ 0 \\ 0 \end{bmatrix} \; ; \; \bar{u} = \begin{bmatrix} \bar{u}_1 \\ \bar{u}_2 \end{bmatrix} = \begin{bmatrix} 0 \\ 0 \end{bmatrix} \quad (26)$$

Then, the states and the inputs are as

$$x = \bar{x} + \delta x \Rightarrow \begin{bmatrix} \beta \\ \omega_r \\ i_c \end{bmatrix} = \begin{bmatrix} \pi/2 + \delta\beta \\ \delta\omega_r \\ \delta i_c \end{bmatrix} \quad (27)$$

$$u = \bar{u} + \delta u \Rightarrow \begin{bmatrix} v_c \\ T_L \end{bmatrix} = \begin{bmatrix} \delta v_c \\ \delta T_L \end{bmatrix} \quad (28)$$

All variables are the same as their deviations except $\beta$, for which new variable $\theta=\delta\beta$ is defined as deviations of angular position around MTPAP. The linearized state-space system is

$$\begin{cases} \dfrac{d}{dt}\delta x(t) = A\,\delta x(t) + B\,\delta u(t) \\ y(t) = C\,\delta x(t) \end{cases} \quad (29)$$

$$A = \begin{bmatrix} \frac{\partial f_1}{\partial x_1} & \frac{\partial f_1}{\partial x_2} & \frac{\partial f_1}{\partial x_3} \\ \frac{\partial f_2}{\partial x_1} & \frac{\partial f_2}{\partial x_2} & \frac{\partial f_2}{\partial x_3} \\ \frac{\partial f_3}{\partial x_1} & \frac{\partial f_3}{\partial x_2} & \frac{\partial f_3}{\partial x_3} \end{bmatrix}, B = \begin{bmatrix} \frac{\partial f_1}{\partial u_1} & \frac{\partial f_1}{\partial u_2} \\ \frac{\partial f_2}{\partial u_1} & \frac{\partial f_2}{\partial u_2} \\ \frac{\partial f_3}{\partial u_1} & \frac{\partial f_3}{\partial u_2} \end{bmatrix} \text{ at } x=\bar{x}, u=\bar{u} \quad (30)$$

It leads to the following linear state-space system

$$\begin{bmatrix} \dot{\theta} \\ \dot{\omega}_r \\ \dot{i}_c \end{bmatrix} = \begin{bmatrix} 0 & 1 & 0 \\ -\frac{K_s}{J} & -\frac{K_d}{J} & \frac{k_t}{J} \\ 0 & -\frac{k_t}{L_{c0}} & -\frac{R}{L_{c0}} \end{bmatrix} \begin{bmatrix} \theta \\ \omega_r \\ i_c \end{bmatrix} + \begin{bmatrix} 0 & 0 \\ 0 & -\frac{1}{J} \\ \frac{1}{L_{c0}} & 0 \end{bmatrix} \begin{bmatrix} v_c \\ T_L \end{bmatrix} \quad (31)$$

The linear electromechanical dynamic is as

$$\begin{cases} v_c = k_t \omega_r + L_c \dfrac{di_c}{dt} + R_c i_c \\ J\dfrac{d^2\theta}{dt^2} + k_d \dfrac{d\theta}{dt} + k_s \theta = k_t i_c - T_L \end{cases} \quad (32)$$

where $k_s=2k_{rest}$. The output is angular position, so $C=[1\;0\;0]^t$.

### D. Transfer Function of Electrical and Mechanical Dynamics

The mechanical dynamics of the actuator is as

$$H_m(s) = \frac{\theta(s)}{I_c(s)} = \frac{k_t}{Js^2 + k_d s + k_s} = \frac{k_t/J}{s^2 + 2\xi\omega_n s + \omega_n^2} \quad (33)$$

where natural frequency and damping ratio are $\omega_n = \sqrt{k_s/J}$ and $\xi = k_d/2J\omega_n$. The electrical dynamic can be written as

$$H'_e(s) = \frac{I_c}{V_c} = \frac{Js^2 + k_d s + k_s}{L_{co}Js^3 + (RJ + L_{co}k_d)s^2 + (Rk_d + k_s k_d + k_t^2)s + Rk_s} \quad (34)$$

where $R$ is the total resistance of coil $R_c$ and current sensor $R_s$. It includes an anti-resonance at the natural frequency of mechanical dynamic. Ignoring the back-emf leaves an RL circuit as

$$H_e(s) = \frac{I_c}{V_c} = \frac{1}{L_{co}s + R} \quad (35)$$

The back-emf is treated as a disturbance in the current loop.

The electrical time constant is $\tau_e \approx L_{c0}/R$.

## V. Eddy-Current Impact on the Electrical Dynamic

To obtain higher accuracy in the electrical dynamic, eddy currents in the laminations and the magnet are modeled, which adds two more degrees of freedom in addition to $L_{c0}$ and $R_c$. As shown in Fig. 5, according to Ampere's law, the stator current $Ni_c$ produces an initial flux $\varphi_0$ whose time variations induce eddy currents in the iron laminations and the magnet ($I_{e.i}$ and $I_{e.m}$) according to Faraday's law which causes a secondary flux attenuating the initial flux. It reduces the coil inductance. A combination of Ampere's and Faraday's laws leads to the diffusion equation $\nabla^2 B = \mu\sigma\, \partial B/\partial t$. To avoid unneeded complexities, the magnet cylinder is simplified to a cube with a rectangular cross-section. The width of the rectangle is the same as the pole width $w_p$. The length of the magnet $l_m$ along the flux loop is obtained such that the cross-sectional areas and thus the volumes are kept the same as

$$w_p l_m = \pi R_r^2 \Rightarrow l_m = \frac{\pi R_r^2}{w_p} \tag{36}$$

The average air-gap length is as

$$l_g = 2 \times \left( \frac{(R_1 - R_r) + (R_2 - R_r)}{2} \right) = R_1 + R_2 - 2R_r \tag{37}$$

The average length of the flux loop within the iron core $l_i$ can be approximated as a half-circle plus pole lengths as

$$l_i = \pi\left(\frac{D_o}{2} - \frac{w_p}{4}\right) + \left\{\left(\frac{D_o}{2} - \frac{w_p}{4}\right) - \left(\frac{l_m}{2} + \frac{l_g}{2}\right)\right\} \tag{38}$$

The reluctances of air-gap, magnet, and iron are obtained as

$$\begin{cases} R_g = \dfrac{l_g}{\mu_0 A_p} \\ R_m = \dfrac{l_m}{\mu_0 A_p} \\ R_i = \dfrac{l_i}{\mu_0 \mu_{ri} A_p} \\ A_p = w_p L \end{cases} \tag{39}$$

where $\mu_{ri}$ is the relative permeability of iron. The area of the left and right return paths, including half of the air-gap flux $\varphi_0/2$, is almost $w_p L/2$. The total reluctance $R_{t0}$ and its approximation based on the low-frequency inductance $L_{c0}$ is as

$$\begin{cases} R_{t0} = R_g + R_m + R_i = \dfrac{l_i + \mu_{ri}(l_g + l_m)}{\mu_0 \mu_{ri} A_p} \\ R_{t0} \approx \dfrac{N^2}{L_{c0}} \end{cases} \tag{40}$$

The initial flux and flux density are obtained as $\varphi_0 = Ni_c/R_{t0}$ and $B_0 = \varphi_0 A_p$. Employing Ampere's law over a flux loop leads to

$$\oint_c \frac{B}{\mu}\cdot dl = I_{enc} \Rightarrow$$

$$\frac{B l_g}{\mu_0} + \frac{B l_m}{\mu_0} + \frac{B l_i}{\mu_0 \mu_{ri}} = Ni_c + I_{e.i} + I_{e.m} \tag{41}$$

It can be rewritten to obtain the effective permeabilities to solve diffusion in the laminations and magnets as

$$\begin{cases} Bl_i = \mu_{eff}^i (NI_c + I_{e.i} + I_{e.m}) \\ \mu_{eff}^i = \dfrac{\mu_0 \mu_{ri} l_i}{l_i + \mu_{ri}(l_g + l_m)} = \dfrac{l_i}{R_t A_p} \approx \dfrac{l_i L_{co}}{N^2 A_p} \end{cases} \quad (42)$$

$$\begin{cases} Bl_m = \mu_{eff}^m (NI_c + I_{e.i} + I_{e.m}) \\ \mu_{eff}^m = \dfrac{\mu_0 \mu_{ri} l_m}{l_i + \mu_{ri}(l_g + l_m)} = \dfrac{l_m}{R_t A_p} \approx \dfrac{l_m L_{co}}{N^2 A_p} \end{cases} \quad (43)$$

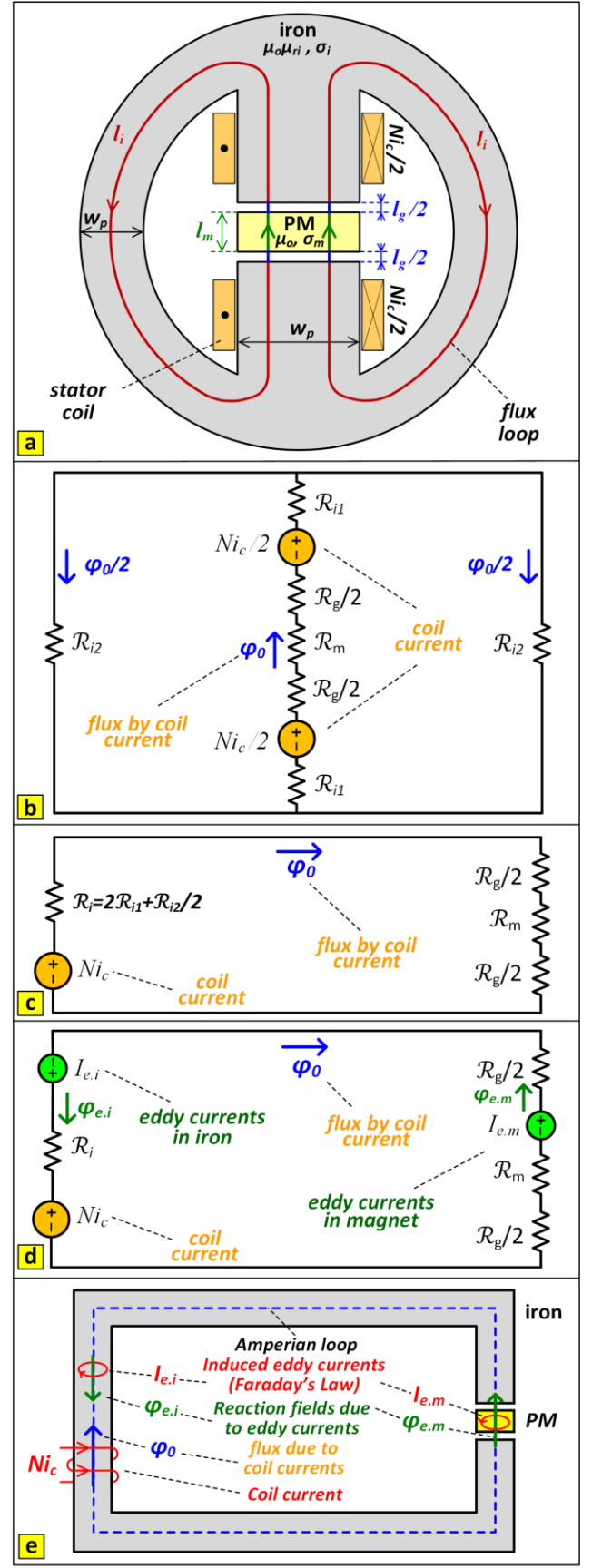

Fig. 5. (a)-(b) MEC and simplified MEC without eddy currents, (c) simplified MEC with eddy currents, and (d) paths of Ampere's and Faraday's laws.

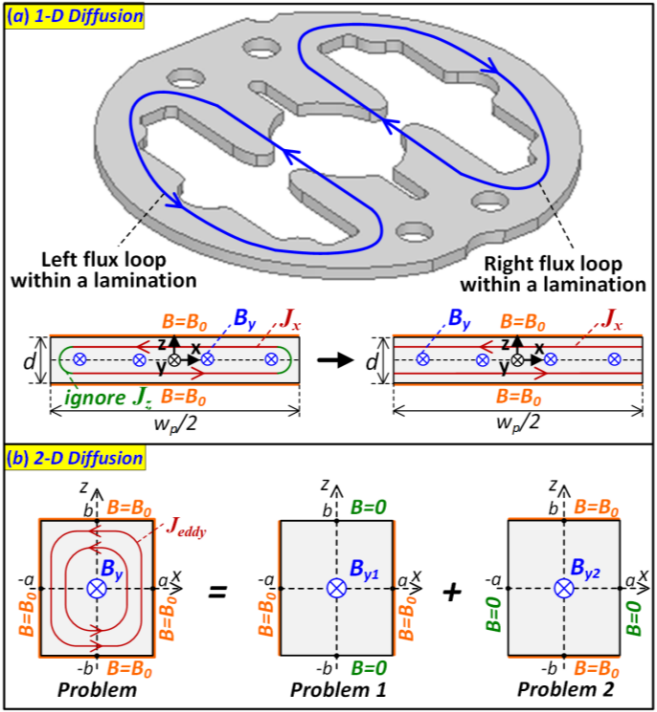

Fig. 6. (a) 1-D diffusion in laminations, and (b) 2-D diffusion in magnet.

### A. 1-D Diffusion for Eddy Currents in the Laminations

As shown in Fig. 6(a), since the laminations are thin, the eddy-currents in the laminations can be modeled by one-dimensional diffusion as

$$\begin{cases} \dfrac{\partial^2 B_y}{\partial z^2} = \mu_{eff}^i \sigma_i \dfrac{\partial B_y}{\partial t} \\ B_y(z,t) = \mathrm{Re}\{\hat{B}_y(z) e^{j\omega t}\} \end{cases} \quad (44)$$

In phasor domain, it leads to

$$\dfrac{\partial^2 \hat{B}_y}{\partial z^2} = j\omega \mu_{eff}^i \sigma_i \hat{B}_y \quad (45)$$

The solution is obtained as

$$s = \pm\sqrt{j\omega\mu_{eff}^i \sigma_i} = \pm\alpha \Rightarrow \\ \hat{B}_y(z) = A_+ e^{+\alpha z} + A_- e^{-\alpha z} \quad (46)$$

As the initial field $B_0$ on the boundaries of the magnet is not disturbed by the flux produced by the eddy currents, the boundary conditions are $\hat{B}_z(x, z = \pm d/2) = B_0$ which result in

$$\begin{bmatrix} e^{\frac{\alpha d}{2}} & e^{-\frac{\alpha d}{2}} \\ e^{-\frac{\alpha d}{2}} & e^{\frac{\alpha d}{2}} \end{bmatrix} \begin{bmatrix} A_+ \\ A_- \end{bmatrix} = \begin{bmatrix} B_0 \\ B_0 \end{bmatrix} \Rightarrow \begin{cases} A_+ = \dfrac{e^{\alpha d/2}}{1 + e^{\alpha d/2}} \\ A_- = \dfrac{e^{\alpha d/2}}{1 + e^{\alpha d/2}} \end{cases} \quad (47)$$

By substituting $A+$ and $A-$, the solution is obtained as

$$\hat{B}_y(z,\omega) = B_0 \dfrac{\cosh \alpha z}{\cosh \dfrac{\alpha d}{2}} \quad (48)$$

The flux passing all laminations is obtained as follows:

$$\begin{cases} \hat{\varphi}(\omega) = 2m \displaystyle\int_{-\frac{w_p}{4}}^{\frac{w_p}{4}} \int_{-\frac{d}{2}}^{\frac{d}{2}} \hat{B}_y(z,\omega) \, dx \, dz = \varphi_0 \dfrac{\tanh \dfrac{\alpha z}{2}}{\dfrac{\alpha d}{2}} \\ \varphi_0 = \dfrac{N i_c}{R_{t0}} \end{cases} \quad (49)$$

where $m$ is the number of laminations such that $L=md$, and 2 is for the two flux loops within the left and the right sides of the stator yoke. Using the approximation $\tanh x = 1/(1+x)$ and substituting for $\varphi_0$, the following MEC is obtained.

$$\begin{cases} \hat{\varphi}(j\omega) = \dfrac{N I_c(j\omega)}{R_{t0} + R_{e.i}(j\omega)} \\ R_{e.i}(j\omega) = R_{t0} \, Q_i(j\omega) \\ Q_i(j\omega) = \dfrac{d}{2}\sqrt{j\omega \mu_{eff}^i \sigma_i} \end{cases} \quad (50)$$

The eddy-impedance $R_{e.i}$ is a half-order complex reluctance that is zero at $\omega=0$. It goes up with frequency, causing a magnitude reduction and a phase lag in the flux $\varphi(t)$ with respect to the magnetomotive force or coil current. The magnetic circuit associated with this equation is shown in Fig. 7(a) with $Ni_c$ as the MMF, and $R_t$ and $R_{e.i}$ as reluctances of the flux path. The induced eddy current density in one lamination is obtained as

$$\begin{cases} \hat{J}(z,\omega) = \dfrac{1}{\mu_{eff}^i} \nabla \times \hat{B}_y \Rightarrow \\ J_x = \dfrac{1}{\mu_{eff}^i} \dfrac{\partial \hat{B}_y}{\partial z} = B_0 \dfrac{\alpha}{\mu_{eff}^i} \dfrac{\sinh \alpha z}{\cosh \dfrac{\alpha d}{2}} \end{cases} \quad (51)$$

### B. 2-D Diffusion for Eddy Currents in the Magnet

As shown in Fig. 6(b), the eddy-currents in the magnet can be modeled using two-dimensional diffusion as

$$\begin{cases} \dfrac{\partial^2 B_y}{\partial x^2} + \dfrac{\partial^2 B_y}{\partial z^2} = \mu_{eff}^m \sigma_m \dfrac{\partial B_y}{\partial t} \\ B_y(x,z,t) = \mathrm{Re}\{\hat{B}_y(x,z) e^{j\omega t}\} \end{cases} \quad (52)$$

where $\hat{B}_y$ is a complex number. In phasor domain, it leads to

$$\dfrac{\partial^2 \hat{B}_y}{\partial x^2} + \dfrac{\partial^2 \hat{B}_y}{\partial z^2} = j\omega \mu_{eff}^m \sigma_m \hat{B}_y \quad (53)$$

Using the separation of variables, we have

$$\hat{B}_y(x,z) = X(x) Z(z) \Rightarrow \dfrac{X''}{X} + \dfrac{Z''}{Z} = j\omega \mu_{eff}^m \sigma_m \quad (54)$$

The boundary conditions are $\hat{B}_y(\pm a, z) = \hat{B}_y(x, \pm b) = B_0$ where $a = w_p/2$, $b = L/2$. By superposition, the problem can be divided into two problems as shown in Fig. 6(b) with boundary conditions

$$P1: \begin{cases} \hat{B}_y(\pm a, z) = B_0 \\ \hat{B}_y(x, z = \pm b) = 0 \end{cases} \quad (55)$$

$$P2: \begin{cases} \hat{B}_y(\pm a, z) = 0 \\ \hat{B}_y(x, z = \pm b) = B_0 \end{cases} \quad (56)$$

The solution of equation (54) for problem 1 is as

$$P1 \begin{cases} \dfrac{X''}{X} = k_{1n}^2 \\ s = \pm k_{1n} \Rightarrow X(x) \sim \sinh k_{1n} x, \cosh k_{1n} x \\ \dfrac{Z''}{Z} = -(\dfrac{n\pi}{2b})^2 \\ s = -j(\dfrac{n\pi}{2b}) \Rightarrow Z(z) \sim \sin \dfrac{n\pi}{2b} z, \cos \dfrac{n\pi}{2b} z \\ k_{1n}^2 - (\dfrac{n\pi}{2b})^2 = j\omega \mu_{\text{eff}}^m \sigma_m \Rightarrow \\ k_{1n} = \sqrt{(\dfrac{n\pi}{2b})^2 + j\omega \mu_{\text{eff}}^m \sigma_m} \end{cases} \quad (57)$$

Satisfying $\hat{B}_y(x, \pm b) = 0$, the solution is obtained as

$$\hat{B}_{y1}(x, z, \omega) = \sum_{n=1,3,\ldots}^{+\infty} a_n \cos \dfrac{n\pi}{2b} z \dfrac{\cosh k_{1n} x}{\cosh k_{1n} a} \quad (58)$$

where $a_n$ is obtained as the coefficients of the Fourier series of the boundary condition $\hat{B}_z(\pm a, z) = B_0$ as

$$a_n = \dfrac{1}{b} \int_{-b}^{b} B_0 \cos \dfrac{n\pi z}{2b} dz = \dfrac{4}{n\pi} \sin \dfrac{n\pi}{2} \quad (59)$$

The solution of equation (48) for problem 2 is as

$$P2 \begin{cases} \dfrac{X''}{X} = -(\dfrac{n\pi}{2a})^2 \\ s = -j(\dfrac{n\pi}{2a}) \Rightarrow X(x) \sim \sin \dfrac{n\pi}{2a} x, \cos \dfrac{n\pi}{2a} x \\ \dfrac{Z''}{Z} = k_{2n}^2 \\ s = \pm k_{2n} \Rightarrow Z(z) \sim \sinh k_{2n} z, \cosh k_{2n} z \\ -(\dfrac{n\pi}{2a})^2 + k_{2n}^2 = j\omega \mu_{\text{eff}}^m \sigma_m \\ k_{2n} = \sqrt{(\dfrac{n\pi}{2a})^2 + j\omega \mu_{\text{eff}}^m \sigma_m} \end{cases} \quad (60)$$

Satisfying $\hat{B}_z(\pm a, 0) = 0$, the solution is obtained as

$$\hat{B}_{y2}(x, z, \omega) = \sum_{n=1,3,\ldots}^{+\infty} b_n \cos \dfrac{n\pi}{2a} x \dfrac{\cosh k_{2n} z}{\cosh k_{2n} b} \quad (61)$$

where $b_n$ is Fourier series coefficients of the boundary condition $\hat{B}_y(\pm a, z) = B_0$ as $b_n = a_n$. Thus $B_y = B_{y1} + B_{y2}$ is obtained as

$$\hat{B}_y = \sum_{n=1,3,\ldots}^{+\infty} \dfrac{4}{n\pi} \sin \dfrac{n\pi}{2} \left( \cos \dfrac{n\pi}{2b} z \dfrac{\cosh k_{1n} x}{\cosh k_{1n} a} + \cos \dfrac{n\pi}{2a} x \dfrac{\cosh k_{2n} z}{\cosh k_{2n} b} \right) \quad (62)$$

By integrating over the PM area, the flux is obtained as

$$\begin{cases} \hat{\varphi}(\omega) = \int_{-b}^{b} \int_{-a}^{a} \hat{B}_y(x, z, \omega) dx dz \\ \hat{\varphi}(\omega) = \sum_{n=1,3,\ldots}^{+\infty} \dfrac{8}{n^2 \pi^2} \varphi_0 \left( \dfrac{\tanh k_{1n} a}{k_{1n} a} + \dfrac{\tanh k_{2n} b}{k_{2n} b} \right) \end{cases} \quad (63)$$

where $\varphi_0 = 4abB_0$. As $a \approx b$, for simplicity of calculations, the rectangle is approximated with a square whose side width $w$ is picked such that the area is the same, i.e. $w = \sqrt{ab}$. Only the fundamental component ($n=1$) is employed to obtain a lumped-element model. The approximation $\tanh x = 1/(1+x)$ is used as well. As the series terms for n=3, 5, ... are ignored, the DC gain should be matched such that $\varphi(\omega=0)=\varphi_0$. Substituting for $\varphi_0 = NI_c/R_{t0}$ and writing the rest in format $1/(1+func(\omega))$ leads to

$$\hat{\varphi}(\omega) = \dfrac{N I_c(j\omega)}{R_{t0} + R_{e.m}(j\omega)} \quad (64)$$

where

$$\begin{cases} R_{e.m}(j\omega) = R_{t0} Q_m(j\omega) \\ Q_m(j\omega) = \dfrac{w\sqrt{\left(\dfrac{\pi}{2w}\right)^2 + j\omega \mu_{\text{eff}}^m \sigma_m} - \dfrac{\pi}{2}}{1 + \dfrac{\pi}{2}} \end{cases} \quad (65)$$

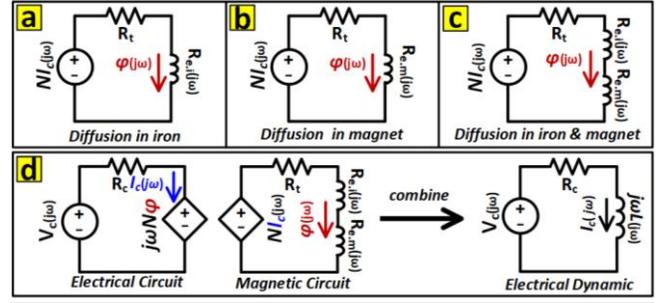

Fig. 7. (a)-(c) MEC with eddy current in iron and magnet, and (d) coupled electric-magnetic circuit to obtain electrical dynamic including eddy currents.

The eddy-impedance $R_{e.m}$ is zero at $\omega=0$. The magnetic circuit associated with this equation is shown in Fig. 7(b) with $NI_c$ as the MMF, and $R_t$ and $R_{e.m}$ as reluctances of the flux path. The induced eddy current density in magnet is as

$$\hat{J}(x, z, \omega) = \dfrac{1}{\mu_{\text{eff}}^m} \nabla \times \hat{B}_y = \dfrac{1}{\mu_{\text{eff}}^m} \left( \dfrac{\partial \hat{B}_y}{\partial z} \hat{a}_x - \dfrac{\partial \hat{B}_y}{\partial x} \hat{a}_z \right) \quad (66)$$

### C. The Coupled Electric-Magnetic Circuit

As shown in Fig. 7(c), the MEC incorporates eddy currents in both laminations and the magnet with a total reluctance of $R_t(j\omega) = R_{t0} + R_{e.i}(j\omega) + R_{e.m}(j\omega)$ and a MMF of $NI_c$. Combining magnetic and electric circuits as in Fig. 7(d) results in the system of equations:

$$\begin{cases} V_c = R_c I_c + j\omega N\varphi \\ NI_c = (R_{t0} + R_{e.i} + R_{e.m})\varphi \end{cases} \Rightarrow \begin{bmatrix} R_c & j\omega N \\ -N & R_{t0} + R_{e.i} + R_{e.m} \end{bmatrix} \begin{bmatrix} I_c \\ \varphi \end{bmatrix} = \begin{bmatrix} V_c \\ 0 \end{bmatrix} \quad (67)$$

It is seen that there is a codependency between electrical and magnetic circuits, where magnetic flux $\varphi$ from the magnetic circuit is returned to the electrical circuit as the dependent voltage source $j\omega N\varphi$, and the electrical current $I_c(j\omega)$ is returned to the magnetic circuit from the electrical circuit as the dependent MMF $NI_c(j\omega)$.

The electrical dynamic can be obtained by solving the above system of equation, or simplifying by finding $\varphi$ from the

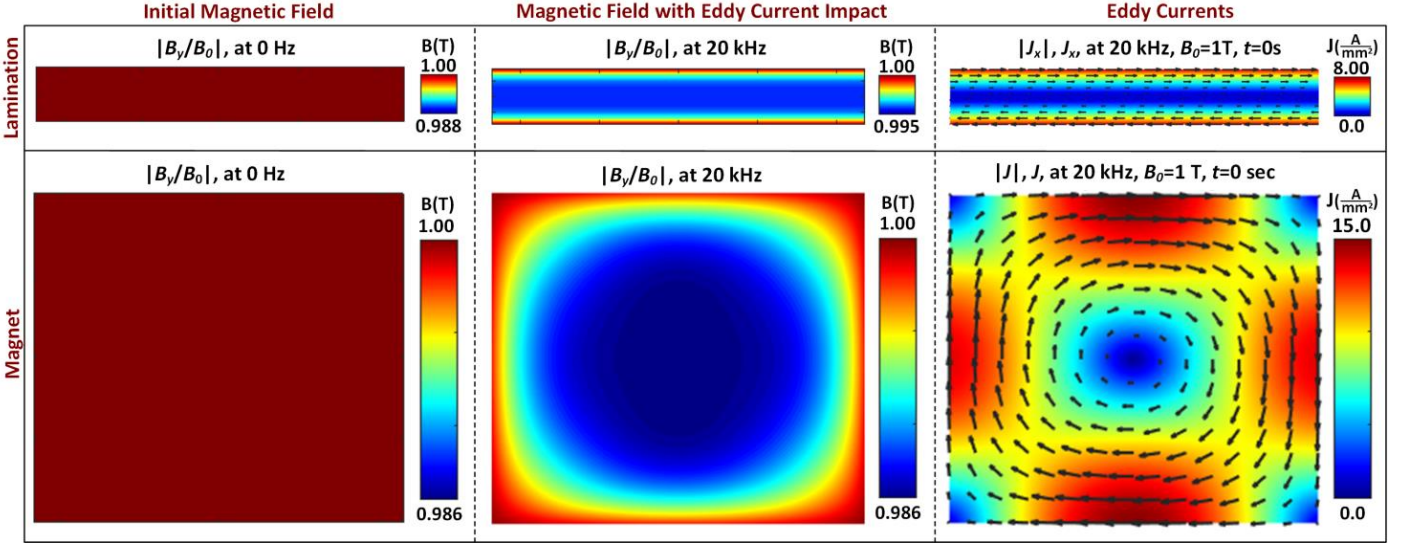

Fig. 8. Flux density distribution, current density distributions and current density vectors within laminations (top) and magnet (bottom).

magnetic equation and substituting it into the electric equation

$$\begin{cases} H_e(j\omega) = \dfrac{1+Q(j\omega)}{R_c + j\omega L_{co} + R_c Q(j\omega)} \\ Q(j\omega) = Q_i(j\omega) + Q_m(j\omega) \end{cases} \quad (68)$$

where $Q(j\omega) \geq 0$. The low-frequency inductance is $L_{c0}=N^2/R_{t0}$ as expected. There are four parameters to be found in identification: $R_c$, $L_{c0}$, $\mu_{eff}^i \sigma_i$ and $\mu_{eff}^m \sigma_m$. The frequency-dependent inductance can also be obtained as

$$\begin{cases} L_c(j\omega) = \dfrac{N^2}{R_t(j\omega)} = \dfrac{N^2}{R_{t0}(1+Q(j\omega))} \\ L_{c0} = \dfrac{N^2}{R_{t0}} \Rightarrow L_c(j\omega) = \dfrac{L_{c0}}{(1+Q(j\omega))} \end{cases} \quad (69)$$

Using the above relationship, we have

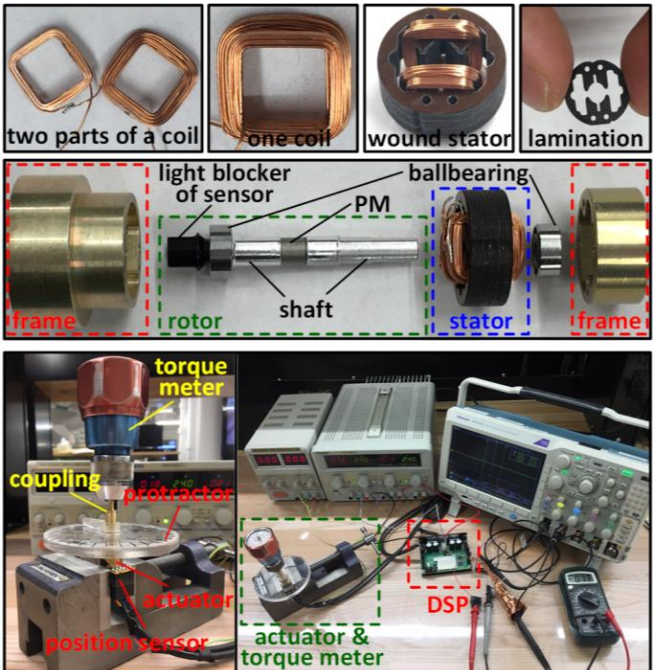

Fig. 9. The prototype actuator (top), and torque-angle measurement (bottom).

$$H_e(j\omega) = \dfrac{1}{R_c + j\omega L_c(j\omega)} \quad (70)$$

Fig. 8 illustrates the distribution of the flux density as well as current density vectors within the laminations and the magnet. It is seen that, at zero frequency, no eddy current is induced and flux density distributions are uniform, while eddy currents are induced at higher frequencies causing a reduction in the flux density at the center of the material.

### D. Fractional-Order System

The square root of $s=j\omega$, illustrate a fractional dynamic which may be written as

$$H_e(s) = \dfrac{\sum_{j=0}^{m} b_j s^{\beta_j}}{\sum_{i=0}^{n} a_i s^{\alpha_i}} \quad (71)$$

where $s^\alpha$ and $s^\beta$ correspond to fractional derivatives. Here $Q_i$ is in the above format, and $Q_m$ can be rewritten using Taylor expansion as

$$\begin{cases} Q_i(s) = \dfrac{d}{2}\sqrt{\mu_{eff}^i \sigma_i}\, s^{\frac{1}{2}} \\ Q_m(s) = \dfrac{w\left(\dfrac{\pi}{2w} + \dfrac{w}{\pi}\mu_{eff}^m \sigma_m s + \dfrac{w^3}{\pi^3}(\mu_{eff}^m \sigma_m)^2 s^2\right) - \dfrac{\pi}{2}}{1 + \dfrac{\pi}{2}} \end{cases} \quad (72)$$

## VI. EXPERIMENTAL EVALUATION AND IDENTIFICATION

Fig. 9 shows the prototyped actuator and the torque-angle measurement setup. Fig. 10 shows the experimental setup including the drive and the control loops.

### A. Torque and Back-EMF Profiles

The torque-angle characteristics at zero coil current (the restoration torque), the coil torque and the total torque at a current of *1A* are given in Fig. 11(a). It should be noted that torque constant is not the peak of the total torque which is only the case in conventional actuators without any reluctance torque. Actually, the torque constant is the peak of the coil torque which cannot be directly measured, and it is obtained by subtracting the restoration torque from the total torque. The torque constant is obtained as $k_t$=1.906 *m N.m/A* by experiment and 1.953 *m N.m/A* by 3D FEM and, i.e., less 2.5% of error. To obtain the restoration torque characteristics, at zero current, the rotor needs to be rotated to capture the reluctance torque at different angles by the torque meter. Then the maximum of the restoration torque is obtained from the peak of the curve as $k_{rest}$=0.318 by experiment and 0.28 by FEM and, i.e., an error of 11%. Then the restoration constant is obtained *as $K_s$=2 $k_{res}$*. Among the sources of the discrepancies are prototyping issues, misalignments, material characteristics, etc. The experimental values are used in the identification. The back-emf waveform at a velocity around *100 rad/sec* is shown in Fig. 11(b), where the peak divided by the velocity is obtained as $k_b$=*1.91 volt.sec/rad* by experiment and *1.96 volt.sec/rad* by FEM and, i.e., an error of less than 3%. The impact of saturation is incorporated in the experimental identification of $k_t$. However, the thickness of the yoke (back iron of stator) is designed to be large enough that the saturation impact is ignorable, which is clear in the very small difference of 0.2% in experimental measurements of torque constant and back-emf constant where $k_t$ =1.906 is measured at nominal current incorporating saturation impact while $k_b$=1.91 is measured at zero current at open-circuit condition of stator.

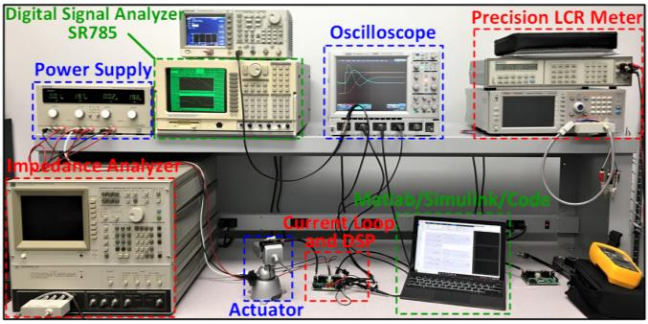

Fig. 10. The setup for identification and analysis of actuator and current loop.

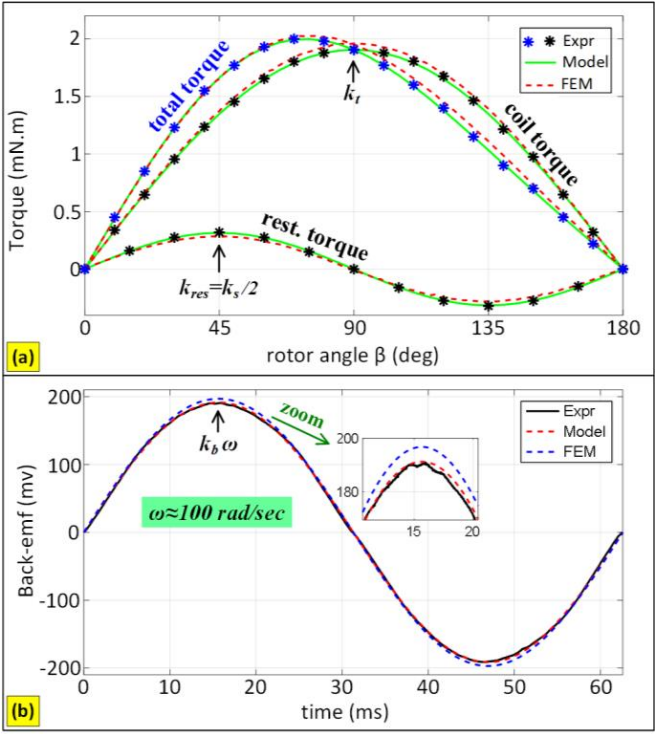

Fig. 11. (a) Coil, restoration and total torques, and (b) back-emf waveform

*B. Identification of the Mechanical Dynamics and Friction*

The identification of the mechanical dynamics can be performed by injecting currents over the desired frequency into the mechanical dynamic. The actuator is excited with the high-bandwidth current control loop as a current source such that the electrical dynamics is eliminated, and thus, the frequency response of the mechanical dynamic $H_m$ can be extracted. In Fig. 12, the waveforms of the coil current $i_c$ and the rotor position $\theta$ as well as frictional hysteresis loops in the torque-position plane for different amplitudes of injected coil current are extracted. The hysteresis loops can be approximated as a straight line whose slope is almost the total stiffness faced by the system. It is observed that, for smaller amplitudes of current, the total stiffness is larger, and the hysteresis band is wider. Fig. 13 shows the frequency response of the mechanical dynamics $H_m$ for different amplitudes of the injected current. A value of 10 mv at the input of the current loop corresponds to a coil current of about 20 mA as the DC gain of the current loop is almost 2. The current amplitudes are also so small so the core is not saturated. It is seen that for very small values of current (e.g. 20 mA) and subsequently torque, the friction dominated and the actuator stops rotating chaotically. Generally, at smaller values of injected current, the current (or torque) profile versus time has more fluctuations due to the higher impact of friction. It can be observed that the DC gain of the mechanical dynamics $H_m$ is smaller than $k_t/k_s$, i.e., the total stiffness of the system is a bit larger than the stiffness of the magnetic spring—surprising! This added stiffness is caused by hysteresis behavior of the pre-sliding friction force which can be described by LuGre model [19]-[20] as follows:

$$F_f = \sigma_s z + \sigma_d \frac{dz}{dt} \tag{73}$$

where $\sigma_s$ is the bristle stiffness, $\sigma_d$ is the bristle damping, $z$ is the friction internal state. We have:

$$\frac{dz}{dt} = v - \frac{\sigma_s |v|}{g(v)} z \tag{74}$$

where $v=d\theta/dt$ is the relative velocity between the two surfaces, and $g(v)$ is the Stribeck curve for steady-state velocities as follows:

$$g(v) = F_c + (F_s - F_c) e^{-(\frac{v}{v_s})^2} \tag{75}$$

where $F_c$ is the Coulomb friction force, $F_s$ is the static friction force, and $v_s$ is Stribeck velocity. A term for viscosity may also be added to $F_f$. Linearization around $z=0$ and $v=0$ results in

$$F_f = \sigma_s \theta + \sigma_d \dot{\theta} \tag{76}$$

In other words, the friction looks like a stiffness $\sigma_s$ and a damping $\sigma_d$. Thus, the mechanical dynamics is then modified to

$$\begin{cases} J\ddot{\theta} + K_d\dot{\theta} + K_s\theta = k_t i_c - F_f \\ \Rightarrow H_m(s) = \dfrac{\theta}{I_c} = \dfrac{k_t}{Js^2 + K_d s + K_s} \end{cases} \quad (77)$$

where $K_d = k_d + \sigma_d$ and $K_s = k_s + \sigma_s$ are the total damping and stiffness of the system. In low frequencies, the frictional hysteresis causes a very small phase delay which gets smaller for larger amplitudes of current as the hysteresis band gets smaller. The profiles of the total stiffness and the low-frequency lag versus current are shown in Fig. 14. The identification is performed using the frequency response for currents around 80-120 mA which is the nominal range of the actuator current, and the values of $K_s$ and $K_d$ do not have big variations. Extracting the DC gain $G_{m0}$ from the magnitude response and $k_t$ from the previous section, the spring factor $K_s = k_t / G_{m0}$ is obtained. At a high-frequency $\omega_{hf}$ where the slope is -40 dB/dec, the inertia dominates the dynamic as $H_m(s) = k_t / Js^2$, so inertia is obtained as $J = k_t / \omega_{hf}^2 | H_m(\omega_{hf}) |$ which is very close to the value obtained by Solid Works. Then, the natural frequency $\omega_n$ can be obtained by $\omega_n = \sqrt{K_s / J}$. Finally, the value of the damping ratio $\zeta$ is set by trial and error until the closest match for the resonance peak of the frequency response is obtained. Then, the damping factor is derived as $K_d = 2J\omega_n\xi$. As shown in Fig. 13, the model shows a good correlation with the experimental results.

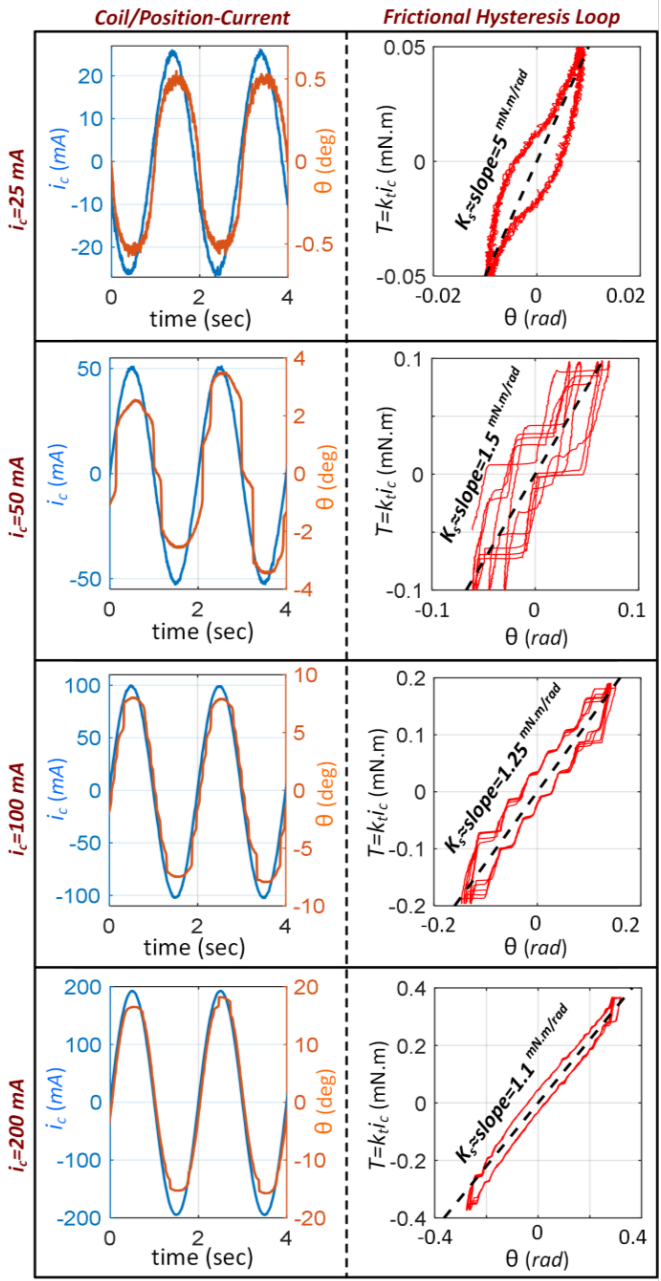

Fig. 12. Mechanical dynamic identification: profiles of the coil current $i_c$ and the position $\theta$ as well as frictional hysteresis loops in the torque-position plane for different amplitudes of injected coil current

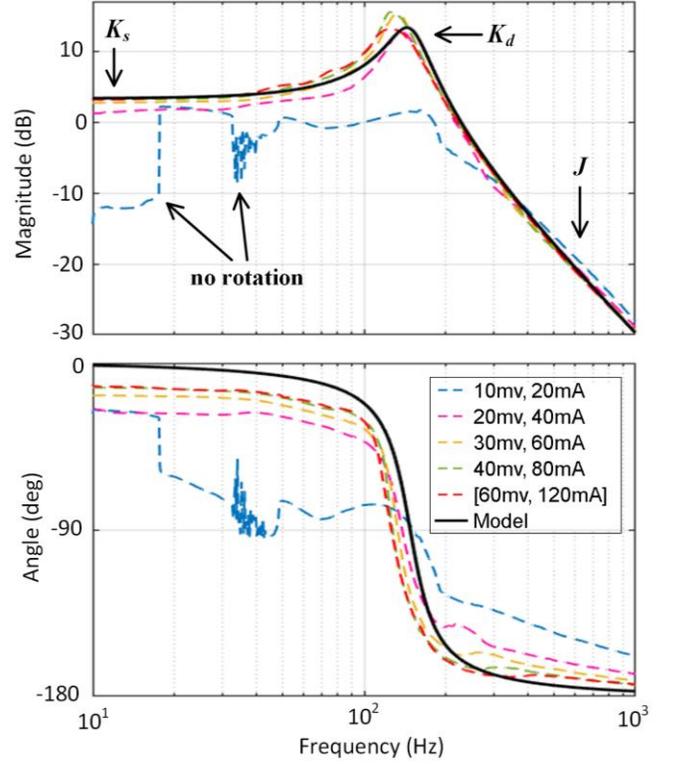

Fig. 13. Mechanical dynamic identification: frequency response of the mechanical dynamics $H_m$ for different amplitudes of injected coil current.

### C. Identification of the Electrical Dynamics

In Fig. 15(a), the 2-DoF conventional RL, the existing 2-DoF eddy current model and the proposed 4-DoF eddy currents models for the electrical dynamics are compared with experimental results. The parameters of the RL model are simply measured by an LCR meter, as given in Table I. From the DC gain, the resistance $R = R_c + R_s$ and then $R_c$ is obtained. At high frequency, the dynamic is reduced to the inductance as $H_e(s) = 1/L_{co}s$, so at a higher frequency $\omega_{hf}$ where the slope is -20 dB/dec, the inductance can be obtained as $L = 1/\omega_{hf} | H_e(\omega_{hf}) |$. These are pretty close to those obtained by LCR meter. The accuracy of this model drops drastically at mid frequencies, causing problems in the design of the current loop and the accuracy of simulation platform. As observed, the phase asymptote of the experimental result, instead of -90°, gets close to -45° due to eddy currents which affect the frequency response by nature of half order (45 degrees).

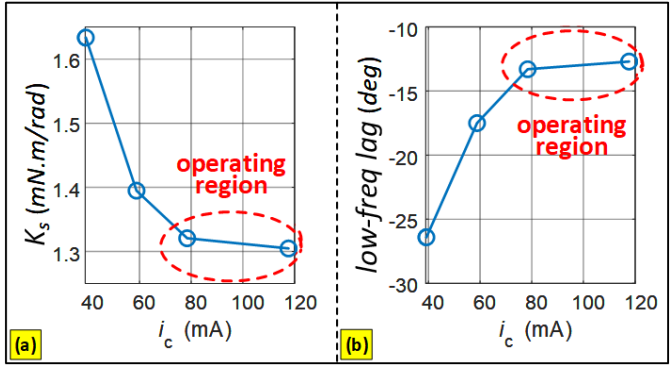

Fig. 14. Mechanical dynamic identification: total stiffness and low-frequency lag due to the hysteresis loop for different amplitudes of injected signal

The phase error at a frequency of 20 kHz (crossover frequency of the current loop) is around 15 degrees in the 2-DoF RL model, while it is reduced to 9 degrees in the 3-DoF model with eddy currents in only laminations [18]-[19], and near-zero 0.4 degrees for the proposed 4-DoF model with eddy currents in both laminations and magnets. The approximated parameters of the 3-DoF model are $R_c$=1.76 Ω, $L_{c0}$=295 μH, $\mu_{eff}^i \sigma_i = 6.4071$. The approximated parameters of the 4-DoF model are $R_c$=1.76 Ω, $L_{c0}$=295 μH, $\mu_{eff}^i \sigma_i = 3.2035$ and $\mu_{eff}^m \sigma_m = 2.8227$. The magnetic reluctances without and with the impact of eddy currents in the laminations and the magnet are shown in Fig. 15(b), illustrating that the reluctance of the flux loop goes up due to eddy currents at higher frequencies, resulting in an inductance reduction.

## VII. CONCLUSION

The paper focuses on the nonlinear and linear modeling of an actuator with magnetic restoration. Analytical calculations of eddy currents in the laminations and the magnet, derived from 1-D and 2-D diffusion equations, along with considerations for pre-sliding friction, contribute to a precise lumped-element model for dynamic studies and control systems designs. Lab experiments using a prototype actuator validate the proposed model, with identification results showing excellent correlation with modeling and FEM. The superiorities of the proposed model over the existing approaches in terms of accuracy and effectiveness is illustrated. In the upcoming Part II of the paper, a drive circuit is proposed, designed, and modeled. A simplified version of the drive circuit is derived for use in the design of the current loop. The paper further delves into illustrating the modeling accuracy and the importance of considering eddy-currents in the design process. Design trade-offs of the drive and the current loop are analyzed, and three position control systems are implemented.

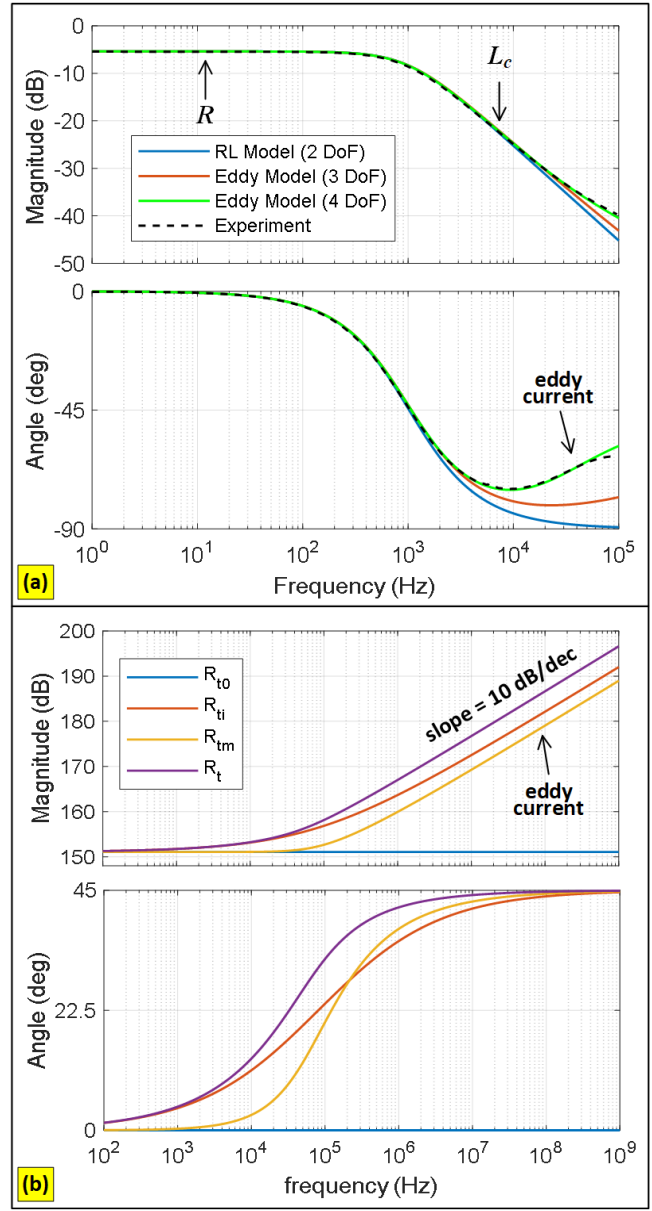

Fig. 15. (a) electrical dynamic and (b) magnetic reluctances.

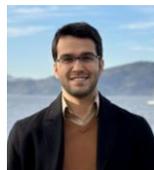

**Sajjad Mohammadi** (S'13) received the B.S. degree in electrical engineering from the Kermanshah University of Technology, Kermanshah, Iran, in 2011, the M.S. degree in electrical engineering from the Amirkabir University of Technology, Tehran, Iran, in 2014, and the M.S. and Ph.D. degrees in electrical engineering and computer science from the Massachusetts Institute of Technology (MIT), Cambridge, MA, USA, in 2019 and 2021, respectively. Now, he is a postdoctoral associate at MIT. His research interests include the design of electric machines, drives, power electronics, and power systems. He attained a number of awards as the 2022 George M. Sprowls Outstanding Ph.D. Thesis Award from MIT and 2014 Best MSc Thesis Award from IEEE Iran Section.

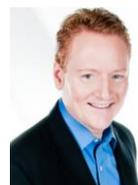

**William R. Benner, Jr.** is president and CTO of Pangolin laser systems. As president, he sets the general strategic direction for the company and oversees all aspects of company operations. As CTO, he is in charge of all hardware and software development as well as research for new products and new directions for the laser display industry. In addition to having received more than 25 international awards for technical achievement, products invented by Benner and manufactured by Pangolin are currently used by some of the best-known entertainment and technology companies in the world, including Walt Disney World, Universal Studios, DreamWorks pictures, Boeing, Samsung, and Lawrence Livermore Labs. Benner holds more than 50 U.S. and International patents.

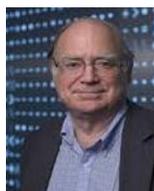

**James L. Kirtley, Jr.** (M'71–SM'80–F'91–LF'11) received the Ph.D. degree from the Massachusetts Institute of Technology (MIT), Cambridge, MA, USA, in 1971. Currently, he is a Professor of electrical engineering at MIT. He was with General Electric, Large Steam Turbine Generator Department, as an Electrical Engineer, for Satcon Technology Corporation as Vice President and General Manager of the Tech Center and as a Chief Scientist and as the Director. He was Gastdozent at the Swiss Federal Institute of Technology, Zurich (ETH), Switzerland. He is a specialist in electric machinery and electric power systems. Prof. Kirtley, Jr. served as Editor-in-Chief of the IEEE TRANSACTIONS ON ENERGY CONVERSION from 1998 to 2006 and continues to serve as an Editor for that journal and as a member of the Editorial Board of the journal Electric Power Components and Systems. He was awarded the IEEE Third Millennium medal in 2000 and the Nikola Tesla prize in 2002. He was elected to the United States National Academy of Engineering in 2007. He is a Registered Professional Engineer in Massachusetts, USA.

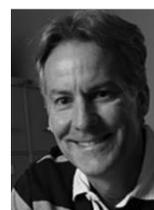

**Jeffrey H. Lang** (F'98) received his SB (1975), SM (1977) and PhD (1980) degrees from the Department of Electrical Engineering and Computer Science at the Massachusetts Institute of Technology. He joined the faculty of MIT in 1980 where he is now the Vitesse Professor of Electrical Engineering. He served as the Associate Director of the MIT Laboratory for Electromagnetic and Electronic Systems from 1991 to 2003, and as an Associate Director of the MIT Microsystems Technology Laboratories from 2012 to 2022. Professor Lang's research and teaching interests focus on the analysis, design and control of electromechanical systems with an emphasis on: rotating machinery; micro/nano-scale (MEMS/NEMS) sensors, actuators and energy converters; flexible structures; and the dual use of electromechanical actuators as motion and force sensors. He has written over 360 papers and holds 36 patents in the areas of electromechanics, MEMS, power electronics and applied control. He has been awarded 6 best-paper prizes from IEEE societies, has received two teaching awards from MIT, and was selected as an MIT MacVicar Fellow in 2022.